\documentclass[preprint,superscriptaddress,showpacs,nofootinbib,showkeys,tightenlines]{revtex4}

\usepackage{amsmath,amsfonts,amssymb,amscd,amsxtra,amsthm}
\usepackage{graphicx}
\usepackage{dcolumn}
\usepackage{bm}
\usepackage{epsfig}

\newcommand{\Slash}[1]{\ooalign{\hfil/\hfil\crcr$#1$}}

\newcommand{\bra}{\langle}
\newcommand{\ket}{\rangle}
\newcommand{\beq}{\begin{eqnarray}}
\newcommand{\eeq}{\end{eqnarray}}

\begin{document}

\title{Non-interacting $KN$ contribution in the QCD sum rule for the pentaquark $\Theta^+(1540)$}

\author{Youngshin Kwon}
\email{kwon@rcnp.osaka-u.ac.jp}
\affiliation{Research Center for Nuclear Physics (RCNP), Osaka University, Ibaraki, Osaka 567-0047, Japan}
\affiliation{Institute of Physics and Applied Physics (IPAP), Yonsei University, Seoul 120-749, Korea}

\author{Atsushi Hosaka}
\email{hosaka@rcnp.osaka-u.ac.jp}
\affiliation{Research Center for Nuclear Physics (RCNP), Osaka University, Ibaraki, Osaka 567-0047, Japan}

\author{Su Houng Lee}
\email{suhoung@phya.yonsei.ac.kr}
\affiliation{Institute of Physics and Applied Physics (IPAP), Yonsei University, Seoul 120-749, Korea}

\begin{abstract}
We perform a QCD sum rule analysis for the pentaquark baryon $\Theta^+$
with the non-interacting $KN$ contribution treated carefully. 
The coupling of the
$\Theta^+$ current to the $KN$ state is evaluated by applying the
soft kaon theorem and vacuum saturation. 
When using a five-quark current including  
scalar and pseudo-scalar diquarks,  the $KN$
contribution turns out not to be very important 
and the previous result of the
negative parity $\Theta^+$ is reproduced again.  
The Borel analysis of the correlation function 
for $\Theta^+$ with the $KN$ continuum states subtracted yields the mass 
of the $J^P= 1/2^-$ $\Theta^+$ around 1.5 GeV.  
\end{abstract}

\pacs{11.30.Er,11.55.Hx,12.38.Aw,14.80.-j}
\keywords{Pentaquark, $\Theta^+$(1540), QCD sum rules}

\maketitle

\section{Introduction}

The experiment by the LEPS Collaboration~\cite{leps} at SPring-8 suggested 
the existence of the exotic baryon $\Theta^+(1540)$ 
of strangeness $S = +1$.  
Subsequent experiments at CLAS collaboration as well as the 
reanalyses of the existing data seem to support the LEPS result.  
Yet, recently many negative results have been also 
reported~\cite{penta04}.  
We need more convincing experiments with high statistics 
especially near the threshold 
regions~\cite{CLASproton}.  

A quantitative analysis was first performed for  exotic baryons 
including $\Theta^+$ 
by Diakonov et. al. using the chiral soliton model~\cite{dpp}.  
They predicted a small mass around 1.5 GeV and a narrow width
of order of ten MeV for $\Theta^+$.  
In the chiral soliton model, $J^P = 1/2^+$ has been also
predicted, with its flavor state being in the antidecuplet 
representation $\bar{10}$ of SU(3).  
Contrary, in a quark model, a five quark state with positive parity 
requires additional orbital excitation and therefore, 
the low mass around 1.5 GeV seems difficult to 
explain~\cite{Oka:2004xh,Takeuchi:2004xm,Huang:2003we}.  
Furthermore, lattice 
QCD~\cite{Sasaki:2003gi,Csikor:2003ng,Takahashi:2005uk} and 
QCD sum rule~\cite{Zhu:2003ba,sdo} analyses
seem to support a negative parity state.  
In view of variation of theoretical predictions, 
dynamics of the low energy QCD is less understood 
than we have thought previously.  

One of the difficulties in the theoretical methods for the exotic 
pentaquark states is that they can couple to scattering 
meson-baryon states.  
In the lattice QCD or QCD sum rule analyses where two point 
correlation functions are studied, 
the five-quark currents for $\Theta^+$ couples to the $KN$ scattering 
state which would contaminate the analysis of the 
resonance of $\Theta^+$.  
This is related to the fact that there are only three color degrees 
of freedom, and therefore hadronic operators of more than 
four quarks must always couple to two (or more) color singlet
hadronic states.  
Therefore, the removal of the $KN$
state is a key issue to understand the properties of $\Theta^+$.  

Motivated by such an observation, Kondo, Morimatsu and Nishikawa
proposed a method to isolate the $KN$ continuum contribution in
their QCD sum rule analysis by Fierz rearranging the five quark
operator into a sum of products of two color-singlet 
components~\cite{kmn}.  
Then they expressed the non-interacting $KN$ contribution as a
convolution of the kaon and nucleon correlation functions. In their
analysis, the $KN$ contribution was large and changed the parity of
$\Theta^+$ before and after the subtraction of the $KN$
contribution. However, in their convolution integral, the momenta
for the kaon and nucleon correlation functions were not always in the
asymptotic region where the OPE can be applicable.

Recently, two of the present authors (S.H.~Lee and Y.~Kwon) and H.~Kim
performed a QCD sum rule analysis where they attempted to isolate
the $KN$ contribution by applying the soft kaon theorem~\cite{lkk}. 
Their
method is reliable concerning the application of the OPE, since it
contains an evaluation of a single correlation function which is
well defined in the asymptotic region. 
It was then found that the $KN$
contribution is negligibly small and the original observation by
Sugiyama, Doi and Oka remained unchanged. 
In their analysis a
coupling of a five quark current to the nucleon state was
considered.
Since the coupling of the ground state nucleon to the five-quark 
current is expected to be small, it naturally 
suppressed the relevant $KN$ contribution. 
Intuitively,
however, the $\Theta^+$ current contains three quark terms which 
is coupled by the ground state nucleon. 
Diagrammatically, these two
processes are depicted in Fig.1. 
Naively, one expects that the
process of Fig.1-(b) would be a major coupling to the $KN$
state. 

In this report, we study the $KN$ scattering contribution in the 
QCD sum rule.  
We apply the previously proposed method of soft kaon theorem
to compute the $KN$ to the vacuum matrix element of the 
pentaquark current.  
This corresponds to the background method when computing 
meson-baryon couplings in the QCD sum rule.  
After taking a commutation relation of the pentaquark current 
with the kaonic axial charge, we find an expression containing 
the scalar operator $\bra q q \ket$.  
This will be replaced by the vacuum condensate by assuming 
the vacuum saturation.  
Such a process can be understood in an effective meson theory 
in terms of the matrix element of the axial vector current 
for the kaon decay.  

\begin{figure}[ht]
\begin{center}
\resizebox{10cm}{!}{\includegraphics{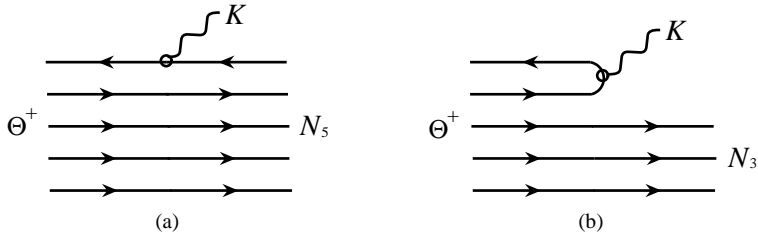}}
\end{center}
\caption{(a) five quark nucleon decay \quad (b) fall apart decay}
\end{figure}

\section{$KN$ contribution to the sum rule}

Let us start with the $KN$ contribution in the two
point correlation function
\begin{equation}
\Pi(q)=\int{{d^4x}\over{(2\pi)^4}}
e^{iq\cdot x}
\langle0|T[J_\Theta(x)\bar{J}_\Theta(0)]|0\rangle \, .  
\label{Piq}
\end{equation}
On the right hand side, the $KN$ scattering state is 
generated by the coupling 
\beq
\langle 0|J_\Theta|KN\rangle
\equiv
i\lambda_{KN}\gamma_5u_N(p) \, , 
\label{lambdaKN}
\eeq
where $u_N(p)$ is a nucleon spinor; 
\begin{equation}
\Pi^{KN}(q)
=
i|\lambda_{KN}|^2
\int{{d^4p}\over{(2\pi)^4}}
{{\gamma_5(\Slash{p}+m_N)\gamma_5}
\over
{p^2-m_N^2}}{{1}\over{(p-q)^2-m_K^2}}\, .
\label{PiKN}
\end{equation}
Hence the determination of the coupling strength 
$\lambda_{KN}$ is the crucial thing.  

To this end, we apply the soft kaon theorem to the 
l.h.s. of Eq.~(\ref{lambdaKN}) 
\begin{equation}
\bra 0|J_\Theta|KN\ket
=
-{{1}\over{f_K}}
\bra 0|[Q^K_5, J_\Theta]|N\ket \, , 
\label{JKN}
\end{equation}
where 
$Q^K_5=\int d^3x\bar{s}\gamma_0\gamma_5u$ 
is the axial charge
with kaon flavor and $f_K$ is the kaon decay constant,
$f_K \sim 100 \mathrm{MeV}$. 
In Ref.~\cite{lkk}, the commutation relation
on the right hand side was computed, leading to the matrix element
$\langle 0|J_{N_5}|N\rangle$ where $J_{N_5}$ is a nucleon current
written in terms of five quarks.

Now consider the matrix element 
$\langle0|J_\Theta|KN\rangle$ 
in a hadronic description. 
We assume that the pentaquark current has a component of $KN$:
\begin{equation}
J_\Theta=aKN+\cdots\, , 
\label{Jexpand}
\end{equation}
where $a$ is a constant, and $K$ and $N$ are kaon and nucleon
fields. 
In terms of quarks, $N$ is written as a three quark current. 
If chiral symmetry is spontaneously broken, the axial charge has 
a piece of time derivative of the Nambu-Goldstone boson field. 
In particular we have
\begin{equation}
Q^K_5=-if_K\int d^3x \, \partial_0 K+\cdots\, .
\label{defQ5}
\end{equation}
Applying the canonical commutation relation 
$[\Pi_K(x), K(y)]=i\delta^3(x-y)$ 
where 
$\Pi_K(x)=\partial_0K(x)$, we find
\begin{equation}
[Q^K_5, J_\Theta]=af_K N + \cdots\, . 
\label{commQ5J}
\end{equation}
In this procedure, the kaon field in (\ref{Jexpand}) has been  
replaced by the decay constant 
$f_K \sim \langle\bar{q}q\rangle$, which is a consequence of
the spontaneous breaking of chiral symmetry. 
In this way we expect
to have a nucleon current written by three quarks as 
in the r.h.s of (\ref{commQ5J}). 

In the QCD sum rule calculation, 
the contributions corresponding to (\ref{commQ5J})
can be computed first extracting the $(\bar{s}q)_K(qqq)_N$ piece by
the Fierz rearrangement of $J_\Theta(\bar{s}qqqq)$, and then 
the $\bar{q}q$
pair in $[Q^K_5, (\bar{s}q)_K]$ is replaced by the vacuum
expectation value.
Let us now consider the following $\Theta^+$ current
\begin{equation}
J_\Theta =
\epsilon^{abc}\epsilon^{def}
\epsilon^{cfg}(u^T_aCd_b)(u^T_dC\gamma_5d_e)C\bar{s}^T_g
\end{equation}
as used in Ref.~\cite{sdo}, 
where $C$ is a charge conjugation matrix
and $T$ denotes transpose. 
As shown in Ref.~\cite{lkk}, we obtain
\begin{equation}
[Q^K_5, J_\Theta]=J_{N_5}\, ,
\label{Q5JT}
\end{equation}
where 
\beq
J_{N_5}
&=&
\epsilon^{abc}
\epsilon^{def}\epsilon^{cfg}
\left[
\{u^T_aC\gamma_5s_b\}\{u^T_dC\gamma_5d_e\}C\bar{s}^T_g
\right.
\nonumber \\
& & 
\left.
+\{u^T_aCd_b\}\{u^T_dCs_e\}C\bar{s}^T_g
+\{u^T_aCd_b\}\{u^T_dC\gamma_5d_e\}C\gamma_5\bar{d}^T_g
\right]\, .
\label{defJN5}
\eeq
Forming a scalar quantity $\bar{q}q$ in (\ref{defJN5}) after the Fierz
rearrangement and replace it by the vacuum expectation value (vacuum
saturation) we obtain
\beq
\langle 0|J_\Theta|KN\rangle
&=&
-{1\over{f_K}}\langle 0|[Q^K_5,
J_\Theta]|N\rangle \nonumber \\
&=&
-{{1}\over{6f_K}}\langle\bar{s}s
+\bar{d}d\rangle\langle0|J_{N_3}|N\rangle
\nonumber \\
&=&-{{f_Km^2_K}\over{6m_s}}\langle0|J_{N_3}|N\rangle\, ,
\label{0JKN}
\eeq
where the three-quark nucleon current is given by
\beq
J_{N_3}=\epsilon^{abc}\left\{ (u^T_aCd_b)u_c
-(u^T_a C\gamma_5 d_b)\gamma_5 u_c\right\} \, .
\label{JN3}
\eeq
In deriving (\ref{0JKN}), 
we have used the Gell-Mann-Oakes-Renner relation:
$f^2_Km^2_K=m_s\langle0|\bar{s}s+\bar{d}d|0\rangle$, 
where $m_s$ is the current mass of the strange quark.  
It is
interesting to see that in Eq.~(\ref{JN3}), 
we find the Ioffe's current
which has the optimal coupling to the physical nucleon state. 
Defining the coupling to the three-quark current by 
\begin{equation}
\langle0|J_{N_3}|N\rangle=i\lambda_{N_3}u_N(q) \, 
\end{equation}
and compare (\ref{0JKN}) with (\ref{lambdaKN}), we find
\beq
\lambda_{KN_3}
=
-{{f_Km^2_K}\over{6m_s}}\lambda_{N_3} \, .
\eeq

\begin{figure}[ht]
\begin{center}
\resizebox{7cm}{!}{\includegraphics{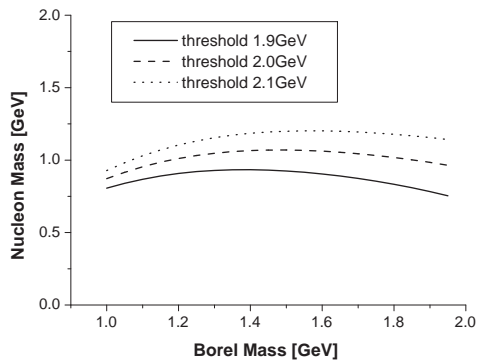}}
\end{center}
\caption{The masses of the nucleon in the QCD sum rule as functions of 
the Borel mass in units of GeV.  Three different values 
1.9, 2.0 and 2.1 GeV are used for the threshold value.}
\end{figure}

In order to determine the coupling $\lambda_{N_3}$, we have
performed the sum rule analysis for the nucleon in an old fashioned
manner with the positive parity states are properly projected out by
using the projector $(1+\gamma_0)/2$. 
Fig.~2 shows the Borel mass analysis for the 
sum rule for nucleon.  
Here we have verified that the Borel stability is 
well achieved.  
From this, we find the mass of the nucleon  
around $1 \mathrm{GeV}$ when a threshold
value $\sqrt{s_0}\sim2.0 \mathrm{GeV}$ is used. 
Furthermore, we find that the coupling
strength 
$|\lambda_{N_3}|^2\simeq1.6\times10^{-4} \mathrm{GeV}^6$. 
Using this value together with 
$f_K \sim 100$ MeV and $m_s \sim 130$ MeV, we obtain
$|\lambda_{KN_3}|^2\simeq14\times10^{-8} \mathrm{GeV}^{10}$, 
which is
larger than the value obtained in Ref.~\cite{lkk} by about 
a factor ten, where 
the coupling of the five-quark current 
$J_{N_5}$ was employed.  
At this point, we have verified that the contribution from the 
three-quark nucleon component is dominant in the coupling 
$\Theta^+ \to KN$.

\section{Reanalysis of the sum rule for $\Theta^+$}

The sum rule for $\Theta^+$ originally performed by 
Sugiyama, Doi and Oka is now modified by subtracting the 
$KN$  contribution.  
It reads 
\begin{equation}
|\lambda_{\Theta\pm}|^{2}e^{-{{m^2_{\Theta}}\over{M^2}}}
=
\int^{\sqrt{s_0}}_0dq_0
e^{-{{q^2_0}\over{M^2}}}\rho^{\pm}_{OPE}(q_0)
-
\int^{\infty}_{m_K+m_N}dq_0
e^{-{{q^2_0}\over{M^2}}}\rho^{\pm}_{KN}(q_0) \, , 
\label{sr_theta}
\end{equation}
where $\rho^{\pm}_{OPE}$ is given in Ref.~\cite{sdo}. 
Because
the l.h.s. is positive definite, the reliable sum rule should have the
r.h.s. with the positive sign also. 
Various contributions of the r.h.s. of Eq.~(\ref{sr_theta}) 
are plotted in Figs.~3 for the positive (right panel) and 
negative (left panel) parity cases using a threshold value 
$\sqrt{s_0}=2.2$ GeV.  
In principle, there remains a question of the convergence 
of the OPE.  
Here we have included, as before, the results up to dimension 6.  
Furthermore we have adopted a threshold value $\sqrt{s_0}=2.2$ GeV, 
slightly larger than the previously used one around 1.8 GeV~\cite{sdo}.  
The larger value seems consistent with the nucleon 
sum rule here.  
Using the larger threshold values, we find that the sum rule 
for the negative parity case is qualitatively similar to the 
previous result~\cite{sdo}, 
while the spectral function of positive parity state 
turns its sign from negative (unphysical) to positive (physical).   
The change in the sign of the positive parity case, however, 
does not necessarily mean the existence of a resonance 
state.  
It indicates that the sum rule for the positive parity 
is rather sensitive to (or equivalently unstable against) 
the choice of the threshold value.   
For the $KN$ contribution, we can see from Figs.~3, 
that it is substantial for the negative parity case, 
while almost negligible for the positive parity case. 
The less important contribution of the positive parity 
may be understood due to the suppression factor of the 
p-wave coupling of the $KN$ state to 
$\Theta^+$~\cite{Hosaka:2004bn}.  

\begin{figure}[ht]
\begin{center}
\resizebox{5.5cm}{!}{\includegraphics{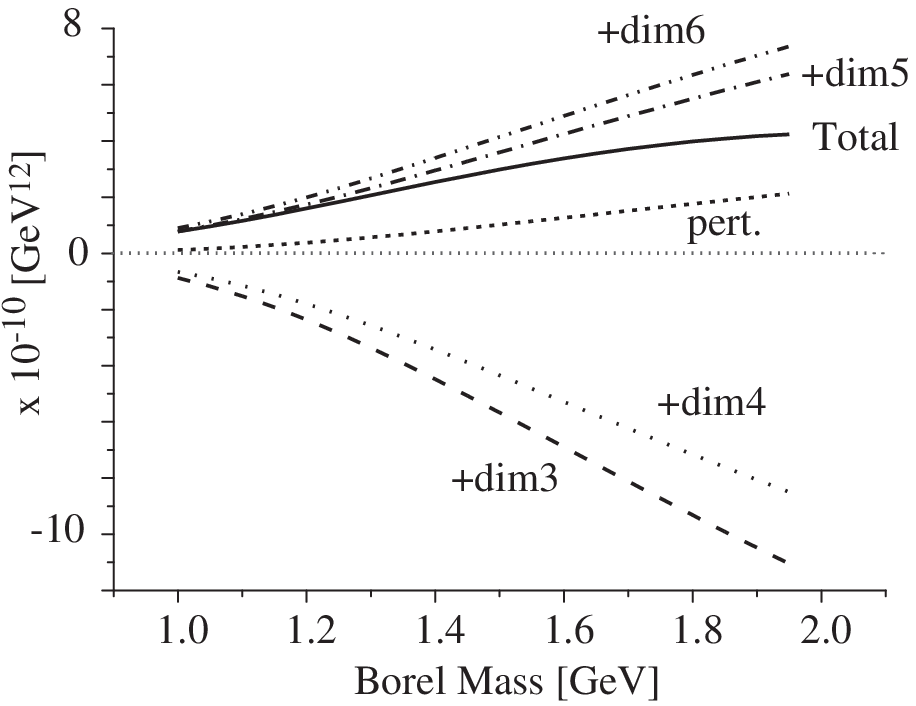}}
\resizebox{5.5cm}{!}{\includegraphics{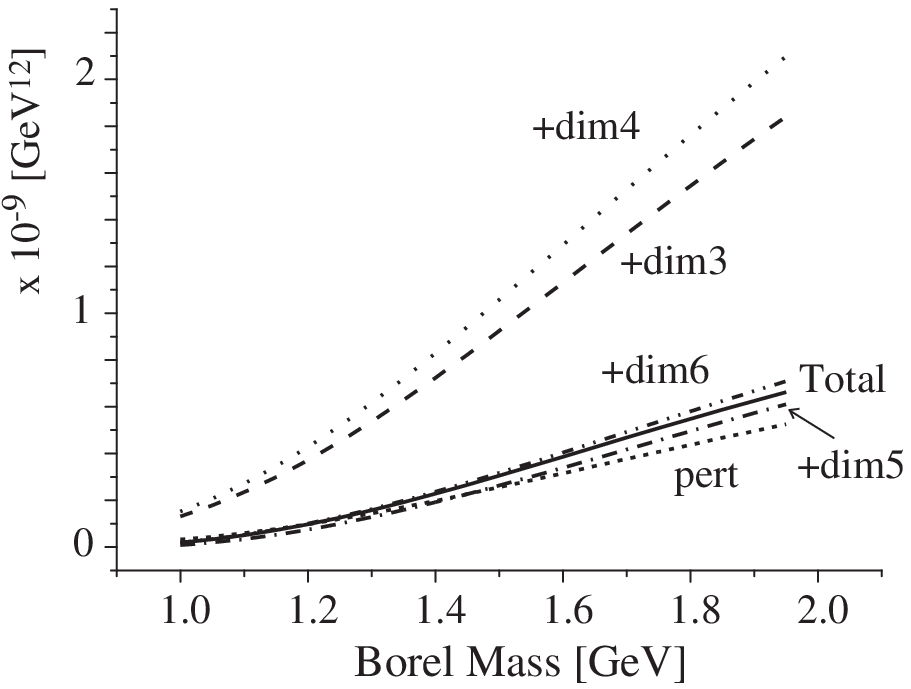}}
\end{center}
\caption{Borel curves for the negative parity (left panel) and
  positive parity (right panel) pentaquarks as we add up the OPE order
  by order. In the total result (solid line) , the $KN$ contribution 
  is subtracted.}
\end{figure}

Now, we have performed the sum rule analysis for 
$\Theta^+$.  
In Fig.~4, we have shown the Borel mass dependence of the 
mass of $\Theta^+$ for the negative parity case 
for three threshold values of 
$\sqrt{s_0}=2.2, 2.3$ and 2.4 GeV.  
Using the larger threshold values, we have found stable 
result as shown in Fig.~4.  
The resulting mass of $\Theta^+$ turns out to be 
around 1.5 -- 1.6 GeV.  
Although not shown here, when 
$\sqrt{s_0}=1.8$ GeV is used, we found unstable result.  
Hence, we have found once again a negative parity $\Theta^+$ 
at the mass region around where experimentally expected.  
On the contrary, for the positive parity case, we could not 
find a realistic solution for the sum rule equation, though  
the l.h.s. of Eq.~(\ref{sr_theta}) is positive.  
As anticipated, this is due to the instability of the spectral function, 
which should not be the case when there is a stable (or quasi-stable)
state.

\begin{figure}[ht]
\begin{center}
\resizebox{7cm}{!}{\includegraphics{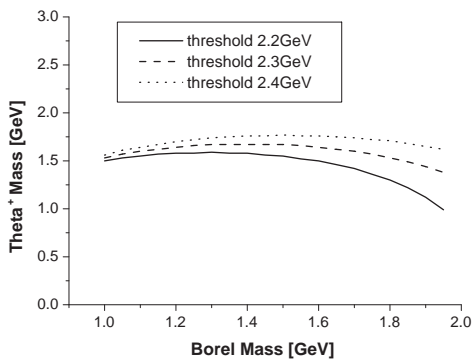}}
\end{center}
\caption{The masses of $\Theta^+$ in the QCD sum rule as functions of 
the Borel mass in units of GeV.  Three different values 
2.2, 2.3 and 2.4 GeV are used for the threshold value.}
\end{figure}

\section{Summary and conclusions}

In this paper we have reanalyzed the QCD sum rule for $\Theta^+$ with
the $KN$ continuum contribution being subtracted out. 
Physically such a contribution should be dominated by the 
coupling of the $\Theta^+$ current to the kaon and three-quark 
nucleon state, which we were able to estimate 
by utilizing the soft kaon theorem and vacuum saturation. 

The $KN$ contribution was found to be substantial for the negative 
parity but almost negligible for the positive parity case.  
Furthermore, we should point out that the spectral 
function of OPE was rather stable 
against the threshold parameter for the negative 
parity case, but extremely sensitive for the positive parity case.  
These results have once again lead to the $\Theta^+$ of 
negative parity with a mass around 1.5-1.6 GeV, while it was not 
possible to find a solution in the positive parity state.  

Although the QCD sum rule is a systematic method based on the 
established technique of QCD, there remain still 
several questions, such as convergence of the OPE and  
dependence of the choice of pentaquark currents.  
Together with the separation of the $KN$ contribution, 
these questions must be very important.  
Further study with proper consideration of them is 
needed for better understanding of the exotic states.


\end{document}